\documentclass[12pt]{article}
\usepackage{mathrsfs}
\usepackage{amsmath}
\usepackage{mathrsfs}
\usepackage{amssymb}
\usepackage{color}
\usepackage{psfrag}
\usepackage{graphicx}
\usepackage{subfigure}
\usepackage{rotating}
\usepackage{cite}
\usepackage[colorlinks=true,linktocpage=true,linkcolor=blue,citecolor=blue]{hyperref}
\usepackage[section]{placeins}
\usepackage{amsfonts}
\usepackage{ifpdf}
\usepackage[toc,page]{appendix}

\newcommand{\bea}{\begin{eqnarray}}
\newcommand{\eea}{\end{eqnarray}}
\newcommand{\be}{\begin{equation}}
\newcommand{\ee}{\end{equation}}

\newcommand{\vs}[1]{\vspace{#1 mm}}

\begin{document}
\topmargin 0pt
\oddsidemargin 0mm

\vspace{2mm}

\begin{center}

{\Large \bf {Phase transition of AdS Schwarzschild black hole and gauge theory dual in presence of external string cloud}}

\vs{10}

{Tanay K. Dey \footnote{E-mail: tanay.dey@gmail.com and tanay.d@smit.smu.edu.in}}

 \vspace{4mm}

{\em
Department of Physics, \\
Sikkim Manipal Institute of Technology,\\
Majitar, Rongpo, East Sikkim, Sikkim-737136,India\\}

\end{center}

\vs{10}

\begin{abstract}
We study the thermodynamics of AdS-Schwarzschild black hole in
the presence of an external string cloud. We observe that, at any temperature, the black
hole configuration is stable with non-zero entropy. We further notice that, when the value of the curvature constant equals to one, if the string cloud
density has  less than a critical value, within a certain range of temperature three black holes configuration exist. One of these black
holes is unstable and other two are stable. At a critical temperature, a transition between these two stable black holes takes place which leads us to conclude that the bound state of quark and anti-quark pairs may not exist. By studying the corresponding dual gauge theory we confirm that the instability of the bound state of quark and anti-quark pair in the dual gauge theory.
\end{abstract}

\newpage
%\tableofcontents

\section{Introduction}\label{intro}
One of the crowning achievements of the Golden Age of Relativity is the discovery that the black holes exhibit thermodynamic properties.
A black hole has a natural temperature associated with its surface gravity and the entropy associated with its area.
These quantities follow classical laws of thermodynamics. In the semi-classical treatment, the black holes
radiate and evaporate eventually. Though the Schwarzschild black hole in an asymptotically flat space-time
has negative specific heat, and is thus thermodynamically unstable, the Schwarzschild black hole in an asymptotically anti-de Sitter
(AdS) space possesses positive specific heat at high temperature and is therefore thermodynamically stable. In their remarkable work
\cite{H-P}, Hawking
and Page further showed that these AdS-Schwarzschild
black holes acquire negative free energy relative to AdS space-time at high temperatures and exhibit a first order phase transition
as one tunes the temperature.
More recently, the study of black holes in AdS space-time has gained a lot of  attention due to Maldacena's discovery of the  AdS/CFT
conjecture \cite{Maldacena}. With in this context, the physics of the black holes or, more precisely, the thermodynamical properties of
the black hole in the bulk AdS space-time play a crucial role in triggering novel behaviour, including phase transitions, of strongly
coupled dual gauge
theories that reside on the boundary of the asymptotically AdS space. This line of investigations started with the work of Witten
who showed that the phase transition that takes place between the thermal AdS at low temperatures and the AdS- Schwarzschild black
hole at high temperatures
could be realized as the confinement/deconfinement transition in the language of boundary $SU(N_c),\, \,{\mathcal N = 4}$ SYM theory
\cite{Witten1, Witten2}. Subsequently, several other extensions of this work appeared. These includes
the consideration of  the R-charges \cite{Chamblin, Myers, Rabin}, addition of the Gauss-Bonnet \cite{Nojiri, Cai,
Cvetic, Cho, Tori, Dey1, Myung1} corrections or the Born-Infeld \cite{Fernando, Dey2, Cai2, Fernando2, Myung} term (
separately and combination \cite{Dey3, Zou, Wei, Zou1} of these terms in the AdS-Schwarzschild black hole) into the action.

There has also been interest to search for the gravity dual of $SU(N_c),\, \, {\mathcal N =
4}$ SYM theory coupled to $N_f$ massless fundamental flavors at finite
temperature and baryon density \cite{Guen, Karch, Head, Big, shankha, Kumar}.
The  fundamental flavors in the dual closed string representation of
$SU(N_c),\, \, {\mathcal N = 4}$ SYM theory corresponds
to adding
open string sector - with one end of the string attached to the boundary of the AdS space and the body
hanged into the bulk and extended up to the center of the AdS space or horizon of the black hole.
In the dual gauge theory, the attached end point of the
string corresponded to the quark or the anti-quark and the body of the string corresponded to the gluonic field of the dual gauge theory. In \cite{Head}, they have also studied the stability of the gravity configurations from the free energy calculation. The free energy is a function of the Polyakov-Maldacena loop(PML). The loop is computed by the area of a certain minimal surface in the dual supergravity background. To fix the (average) area of the appropriate minimal surface they introduce a Lagrange multiplier term into the bulk action. This term, which can also be viewed as a chemical potential for the PML, contributes to the bulk stress tensor like a string stretching from the horizon to the boundary. They find the corresponding “hedgehog” black hole solutions numerically, within the
 $SO(6)$ preserving ansatz.

Motivated by these developments, in this paper we first study the thermodynamics of the recently developed AdS-Schwarzschild black hole
in presence of an external
string cloud \cite{shankha}. In this work they have considered the gravity action of the AdS-Schwarzschild space time with the contribution of the external matter which comprises of  uniformly distributed
strings, each of whose one end is stuck on the boundary. We observe that the black hole configuration is a stable one at any
temperature compared to the AdS
configuration. Even at zero temperature, there is a black hole with a minimum radius. The size crucially depends on the density
of
the string cloud. We see that the density of this cloud plays not only an important role
in finding  out the minimum radius of black hole, it is also an important parameter controlling
the number of black holes present at any given temperature. If the cloud density is greater than a critical value,
there exists only one
black hole. While for the cloud density less than the critical value, and for the value of curvature constant one, within a certain range
of temperature there exist three black holes. Beyond this temperature range again we have a single black hole configuration. Depending on
their sizes, we call them small, medium and large black holes. Among these three holes, the small and the large come with positive
specific
heat, and, the remaining one has a negative specific heat. Therefore, except the medium one, the other two can be stable. Due to the presence of large number of strings the AdS background is deformed to a small black hole background.  We study the
stability of these two black holes by analyzing their free energies as a function of the temperature and the Landau function as a
function of  their radii (at different temperatures). Our observation is that there is a critical temperature below which the small
black hole is the stable configuration and above the critical temperature large black hole is the stable one. At the critical
temperature a transition between the small and the large black holes takes place. Though our approach is different but in the context of stability of the space time configuration our result is almost same as \cite{Head}. Therefore, we suspect that this may lead to an instability of the bound states
of quark and anti-quark pairs in the dual gauge theory. In order to test our suspect finally we study the dual gauge theory along the line of \cite{Yang, Yuan}.

 In the dual gauge theory we consider a probe string whose end points are attached on the boundary of the black hole background and the body of the string is hanged in to the bulk space-time. In the bulk space-time, there are two configurations of the open string. One is the U-shape configuration, where the body of the string reach up to a maximum distance from the boundary. The other one is the straight configuration where the open string reach up to the horizon of the black hole. The first configuration corresponds to the confined state of quark and anti-quark pairs and the second configuration corresponds to deconfined state of quark and anti-quark pairs. We see that for any black hole back ground, an open string is in the U-shape configuration for short distance between quark and anti-quark pair and is in the straight shape configuration for large distance. Thus black hole configuration of the gravity corresponds to a deconfined state of quark and anti-quark pairs in the dual gauge theory.

We have organized our paper as follows:  we start by writing the action of the AdS-Schwarzschild black hole with the matter contribution
coming from the infinitely long string and the corresponding black hole solution in section \ref{gravitydual}. Then we compute the
thermodynamical quantities in section \ref{thermo}. Section \ref{phase} is devoted to the study the different phases of the black
holes. Before summarising our work we study the dual gauge theory in section \ref{gaugetheory}. Finally we summarize in section \ref{conclude}.

\section{ AdS-Schwarzschild black hole in presence of external string cloud}\label{gravitydual}
We start this section by considering the $(n + 1)$ dimensional gravitational action in presence of cosmological constant
with the contribution of the external string cloud,
\begin{equation}
\mathcal{S} = \frac{1}{16 \pi G_{n+1}} \int dx^{n+1} {\sqrt {-g}}( R - 2 \Lambda ) + S_m,
\label{totac}
\end{equation}
here $S_m$ represents the contribution of the string cloud and can be expressed by the following way;
\begin{equation}
S_m = -{\frac{1}{2}} \sum_i {\cal{T}}_i \int d^2\xi {\sqrt{-h}}
h^{\alpha \beta} \partial_\alpha X^\mu \partial_\beta X^\nu g_{\mu\nu},
\label{matac}
\end{equation}
where $g^{\mu\nu}$ and $h^{\alpha \beta}$ are the space-time
and world-sheet metric respectively with $\mu, \nu$ represents
space-time directions and $\alpha, \beta$ stands for world sheet coordinates. $S_m$ is a sum over all the string contributions and ${\cal{T}}_i$ is the tension of $i$'th string. The integration in  ({\ref{matac}}) is taken over the two dimensional string coordinates.
The action (\ref{totac}) possesses black hole solutions and the metric solution of this black hole can be written as
\begin{equation}
ds^2  = -g_{tt}(r) dt^2 + g_{rr}(r) dr^2 + r^2 g_{ij} dx^i dx^j.
\label{genmet}
\end{equation}
Here $g_{ij}$ is the metric on the $(n-1)$ dimensional boundary and
\begin{equation}
g_{tt}(r) = K +\frac{r^2}{l^2} - \frac{2 m}{r^{n-2}} - \frac{2 a}{(n-1) r^{n-3}}= \frac{1}{g_{rr}},
\label{comp}
\end{equation}
where $K = 0, 1, -1$ depending on whether the $(n-1)$ dimensional boundary is
flat, spherical or hyperbolic respectively,  having the  boundary curvature $(n-1)(n-2) K$ and volume
$V_{n-1}$. The uniformly distributed string cloud density $a$ can be written as
\begin{equation}
a(x) = T \sum_i  \delta_i^{(n-1)}(x - X_i),~~~{\rm with} ~a > 0.
\label{denfun}
\end{equation}
In writing $g_{tt}(r)$, the cosmological constant is parameterized  as $\Lambda = - n(n-1)/(2 l^2)$. With equation (\ref{comp}), the metric (\ref{genmet}) represents a black hole with singularity at $r=0$ and the horizon is located at $g_{tt}(r) =0$. The horizon radius, denoted by $r_+$, satisfies the equation
\begin{equation}
 K +\frac{r_+^2}{l^2} - \frac{2 m}{r_+^{n-2}} - \frac{2 a}{(n-1) r_+^{n-3}} = 0.
\label{hor}
\end{equation}
This allows us to write the integration constant $m$ in terms of horizon radius as follows
\begin{equation}
m = K\frac{r_+^{n-2}}{2}+\frac{(n-1) r_+^n - 2 a l^2 r_+}{2 (n-1) l^2}.
\label{mas}
\end{equation}
The integration constant $m$ is related to the ADM ($M$) mass of the black hole as,
\begin{equation}
M = \frac{(n-1) V_{n-1} m}{8 \pi G_{n+1}}.
\end{equation}
Therefore the mass of the black hole can finally be written in the following form
\begin{equation}
M =\frac{(n-1) V_{n-1} }{8 \pi G_{n+1}} \Big[K\frac{r_+^{n-2}}{2}+\frac{(n-1) r_+^n - 2 a l^2 r_+}{2 (n-1) l^2}\Big].
\label{mas}
\end{equation}

Analysing the black hole metric solution and mass, in the next sections we discuss the thermodynamics of this type of black holes. We therefore first compute the thermodynamical quantities.

\section{Thermodynamical quantities}\label{thermo}
It has been well understood that black holes behave as thermodynamic systems. The laws of black hole mechanics become similar to the usual laws of thermodynamics after appropriate identifications between the black hole parameters and the thermodynamical variables.  In order to study the thermodynamics of black holes we first come across various thermodynamical quantities as calculated in the following portions;
Firstly, the temperature of the black holes is found by the following standard formula;
\begin{equation}
T = \frac{1}{4 \pi}\frac{d g_{tt}}{dr}|_{r = r_+}
= \frac{ n(n-1)r_+^{n+2} + K(n-1)(n-2)l^2r_+^n - 2 a l^2 r_+^3}{4 \pi (n-1) l^2 r_+^{n+1}}.
\label{flattemp}
\end{equation}
To find out the entropy we expect that these black holes satisfy the first law of thermodynamics. Therefore by using the first law of thermodynamics, we calculate the entropy and it takes the form as;
\begin{equation}
S = \int T^{-1} dM,
\end{equation}
which satisfies the universal area law of the entropy,
\begin{equation}
S = \frac{V_{n-1}r_+^{n-1}}{4 G_{n+1}}.
\end{equation}
Then we compute the specific heat associated with the black holes and is found as;
\begin{equation}
C = \frac{\partial M}{\partial T} =
\frac{(n-1)V_{n-1}}{4 G_{n+1}}\Big[\frac{n(n-1)r_+^{2n-2}+K(n-1)(n -2)l^2 r_+^{2n-4} - 2 a l^2 r_+^{n-1}}{n(n-1) r_+^{n-1} -K(n-1)(n-2)l^2r_+^{n-3} +
2(n-2) a l^2}\Big].
\label{sh}
\end{equation}
Free energy can be calculated by using the formula
\begin{equation}\label{fenergy}
F=E-TS= \frac{V_{n-1}}{16\pi G_{n+1}}\Big[K r_+^{n-2}-\frac{r_+^n}{l^2}-\frac{(n-2)2a r_+}{(n-1)}\Big].
\end{equation}
Where $E$ is the energy of the black hole which is considered equal to the mass of the black hole.

Finally we also compute the Landau function around the critical point by considering the radius of the black hole as an order parameter to get the better understanding on the phase structure. The Landau function depends on the order parameter $r_+$ and temperature $T$  in the following way,
\begin{equation}\label{landau}
G=\frac{V_{n-1}}{16\pi l^2G_{n+1}}\Big[(n-1)r_+^n -4\pi l^2 T r_+^{n-1}+ K (n-1)l^2r_+^{n-2}-2 a l^2 r_+\Big].
\end{equation}
At the extreme point of this function that is when $\frac{\partial G}{\partial r_+} =0$, we get back the expression
of the temperature given in (\ref{flattemp}). Also, if we plug in the expression of
temperature in to (\ref{landau}), $G$ reduces to the free energy given in (\ref{fenergy}).

Many interesting features of these black holes, related to local and global stabilities,
can be studied  from the detailed analysis of the thermodynamic quantities. We study the thermodynamical phases of these kind of black holes in the next section.

\section{Phases of black hole}\label{phase}
\begin{figure}[h]
\begin{center}
\begin{psfrags}
\psfrag{T}{$\bar T$}
\psfrag{r}{$\bar r$}
\mbox{\subfigure[]{\includegraphics[width=6.5 cm]{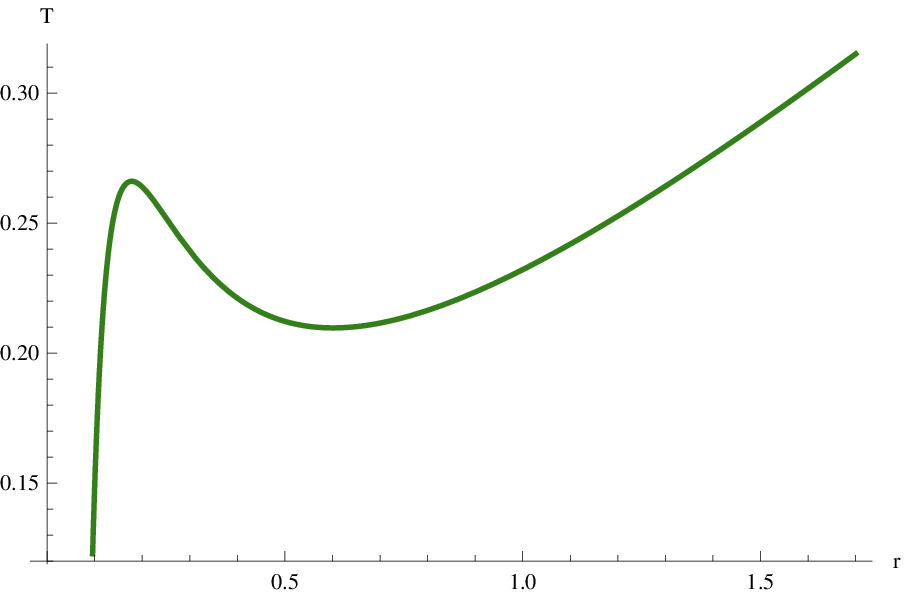}}
\quad
\subfigure[]{\includegraphics[width=6.5 cm]{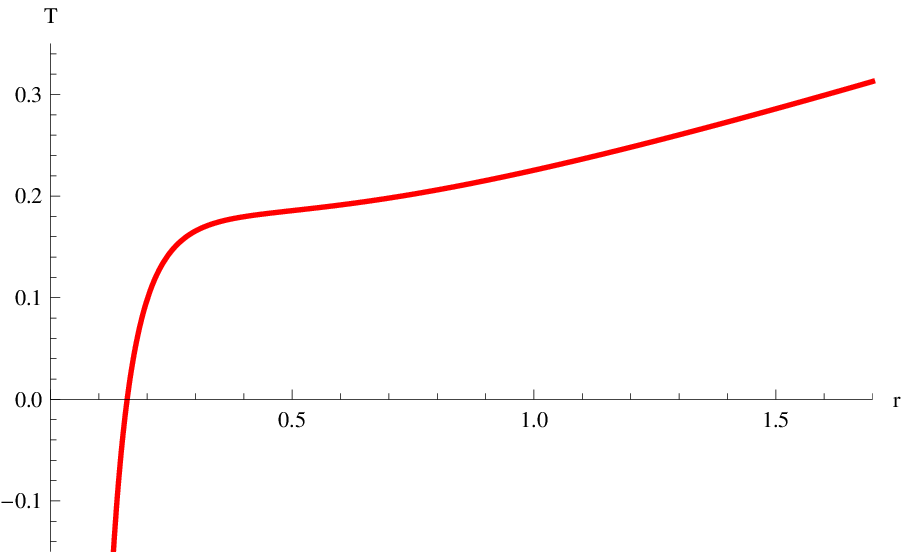}}}
\end{psfrags}
\caption{The plot (a) is for $\bar a=0.3 < \bar a_c, K=1 \,{\rm and}\, l=2$ and plot (b) is for $\bar a=0.5 > \bar a_c, K=1 \,{\rm and}\, l=2$.
}
\label{temp}
\end{center}
\end{figure}

In this section we consider the black holes in five dimensions $(n = 4)$ and the results can  easily be extrapolated in the higher dimensions. We start the study by considering two dimensionless quantities  $\bar a = \frac{a}{l}$  and $ \bar r = \frac{r_+}{l}$ . In terms of these dimensionless quantities the temperature can be expressed in the following form;
\begin{equation}
\bar T=\frac{1}{6\pi l \bar r^2}\big[6\bar r^3 + 3 K \bar r - \bar a\big].
\end{equation}
The behaviour of temperature with respect to $\bar r$  for string cloud density $\bar a$ less or greater than a critical value $\bar a_c$  and  $ K =1$ are drawn in figure \ref{temp}.
For $\bar a   < \bar a_c$, we notice that even at zero temperature black hole exists. The size of the zero temperature black hole can be found in terms of power series of  $\bar a$ which takes the form
\begin{equation}
\bar r_{0} = \frac{\bar a}{3}  + \mathcal{O}(\bar a^2).
\label{rmin}
\end{equation}
Further from figure \ref{temp}(a), it is also observed that at low temperature only one black hole exist. As temperature increases the size of this black hole slowly increases and at a critical temperature with the existing black hole, two new black holes nucleate. Radius of one of these two new black holes reduces and the other one increases when temperature increases. Depending on the size of these three black holes we call them small, medium and large. Up to a certain value of temperature all these three black holes exist and after that small and medium size black holes merge together and vanish. Finally, only large black hole exists at high temperature.

For $\bar a > \bar a_c$, the figure shows that at any temperature only one black hole exists. However when the size of the black hole becomes small, the associated temperature is negative. To avoid the negative temperature of the black hole, the radius of the black hole should be protected by a minimum size which is equal to $ \bar r_{0}$. Therefore for any value of string cloud density there will be a black hole of finite size with non-zero entropy.

\begin{figure}[h]
\begin{center}
\begin{psfrags}
\psfrag{CP}{$\bar C$}
\psfrag{r}[][]{$\bar r$}
\mbox{\subfigure[]{\includegraphics[width=6.5 cm]{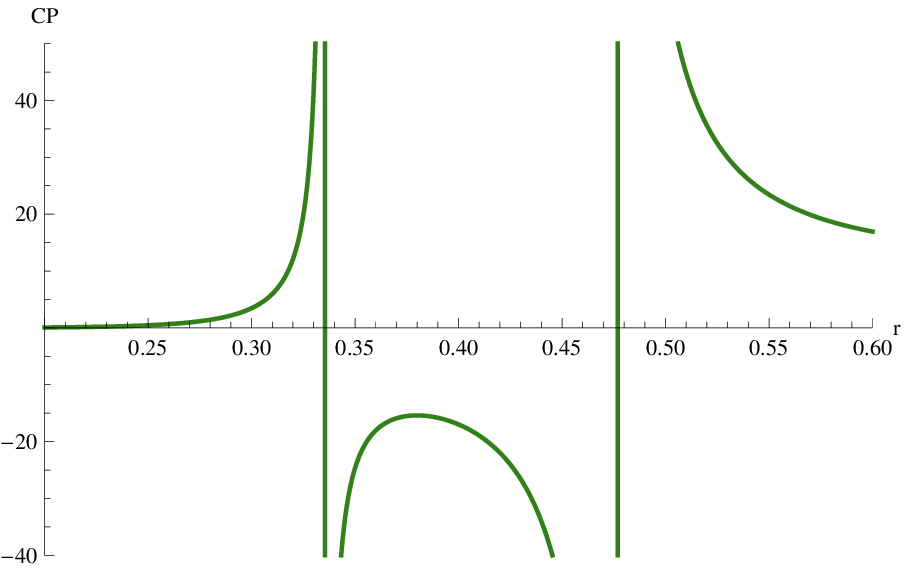}}
\quad
\subfigure[]{\includegraphics[width=6.5 cm]{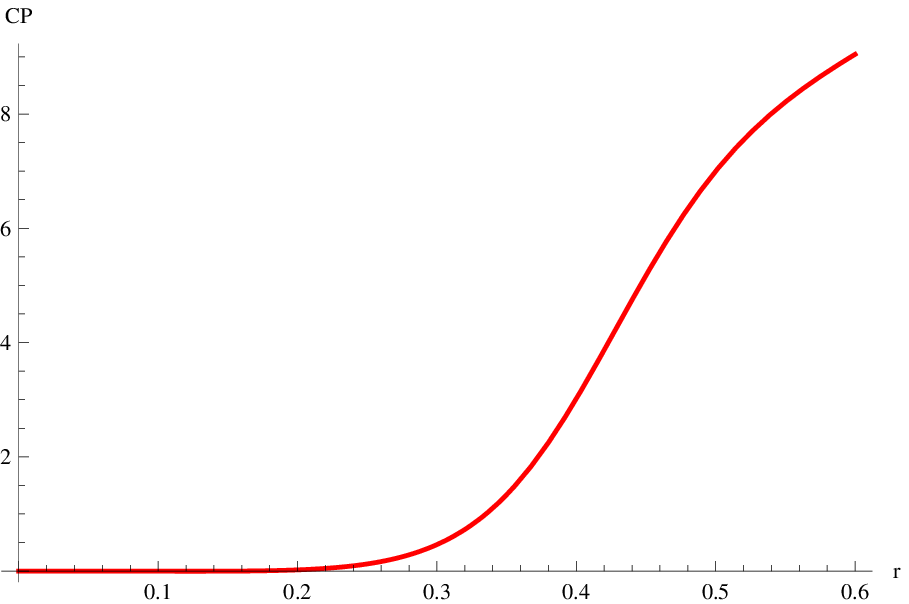}}}
\end{psfrags}
\caption{The plot (a) is for $\bar a=0.39 < \bar a_c, K=1 \,{\rm and}\, l=2$ and plot (b) is for $\bar a=0.5 > \bar a_c, K=1 \,{\rm and}\, l=2$.}
\label{specific}
\end{center}
\end{figure}

From the above discussion it is clear that the space time configuration crucially depends on the critical value of string cloud density $\bar a_c$. Later on we see that the behaviour of thermodynamical quantities also depends on the $\bar a_c$.  Therefore we should find out the critical value $\bar a_c$. To calculate it, we focus on figure \ref{temp}. If $\bar a$ is above the critical value $\bar a_c$, temperature is monotonically increasing function of radius of black hole. However, below the critical value, temperature has two extrema for two real values of black hole radius. So, $\frac{\partial T}{\partial \bar r} =0$, should give two real values of radius. To get the two real values, we find that the string cloud density $\bar a$ has to be less than the critical value;
 $$\bar a_c =\frac{1}{\sqrt 6}\approx 0.408$$
In order to study the stability of these black holes we study the specific heat associated with them. The specific heat by considering the above dimensionless quantities can be written as;
\begin{equation}
\bar C=\frac{3 l^3 V_3}{4 G_5}\Big[\frac{6\bar r^6 + 3 K \bar r^4 - \bar a \bar r^3}{6\bar r^6 - 3 K \bar r + 2\bar a}\Big]
\end{equation}
We study the specific heat as per the figure \ref{specific} where the specific heat is plotted as a function of $\bar r$.
From the figure \ref{specific}, it becomes evident that for $\bar a   < \bar a_c$, the specific heat is positive for small and large sized black hole, while it is negative for the medium sized black hole. Therefore it can be expected that the black holes with positive specific heat can be stable while the black hole with negative specific heat is unstable. For $\bar a > \bar a_c$, we notice that the specific heat monotonically increases from zero value with the increase in the radius of the black hole. So this black hole can also be stable.

We then analyse the free energy to check the stability further. The free energy in terms of dimensionless quantity $ \bar r$ can be rewritten as;
\begin{equation}
\bar F=\frac{V_3 l^2 \bar r}{16\pi G_5}\big[K \bar r -\bar r^3- \frac{4}{3}\bar a\big]
\end{equation}
\begin{figure}[h]
\begin{center}
\begin{psfrags}
\psfrag{F}[][]{$\bar F$}
\psfrag{r}[][]{$\bar r$}
\mbox{\subfigure[]{\includegraphics[width=6.5 cm]{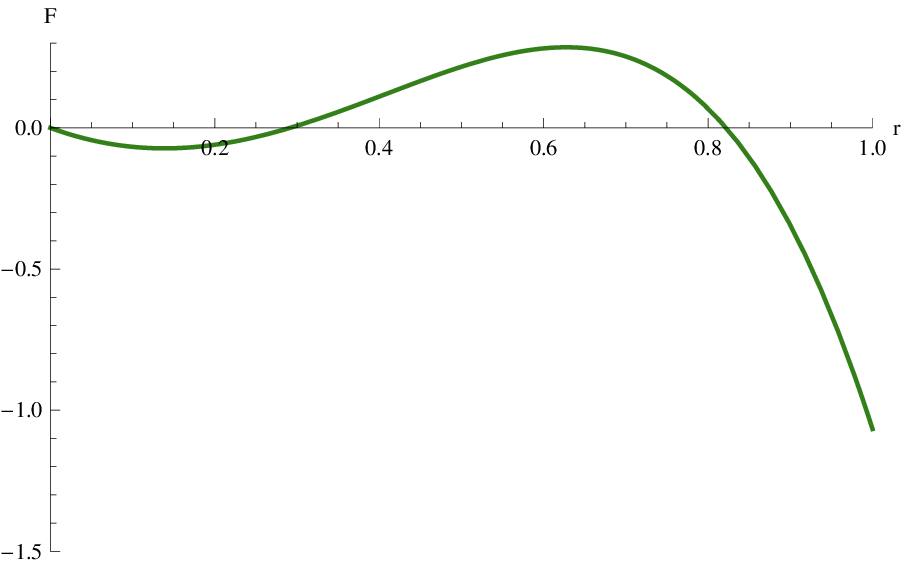}}
\quad
\subfigure[]{\includegraphics[width=6.5 cm]{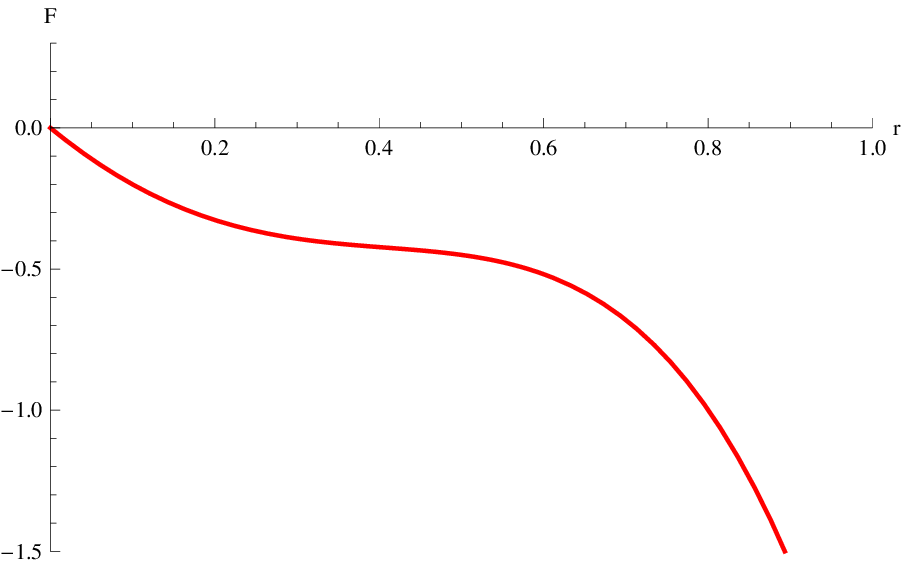}}}
\end{psfrags}
\caption{The plot (a) is for $\bar a=0.2 < \bar a_c, K=1 \,{\rm and}\, l=2$ and plot (b) is for $\bar a=0.45 > \bar a_c, K=1 \,{\rm and}\, l=2$.}
\label{freeenergy}
\end{center}
\end{figure}
The figure \ref{freeenergy} shows that for $\bar a < \bar a_c$, the free energy starts from zero value at $\bar r  = 0 $ and increases towards the negative value with the increase of black hole radius. At a certain value of radius, free energy reaches to the minimum value and then goes to the maximum value with the increase of radius. Again it drops down to the negative region and continues to increase towards the negative value with the increase of radius. Therefore the first extrema which corresponds to small size black hole will be preferable configuration compared to the AdS configuration since its free energy is less than the latter one. However, the free energy of the small black hole is greater than the large size black hole. So the large size black hole should be more stable compared to the smaller one and there is a possibility of having a phase transition between these two black holes.  For $\bar a > \bar a_c$, again the free energy monotonically decreases with the radius. So the black hole configuration is the stable one.

Now to verify the possibility of the phase transition between these black holes we study the free energy in terms of temperature.
\begin{figure}[h]
\begin{center}
\begin{psfrags}
\psfrag{F}{$\bar F$}
\psfrag{T}{$\bar T$}
\mbox{\subfigure[]{\includegraphics[height=6.0cm,width=6.5 cm]{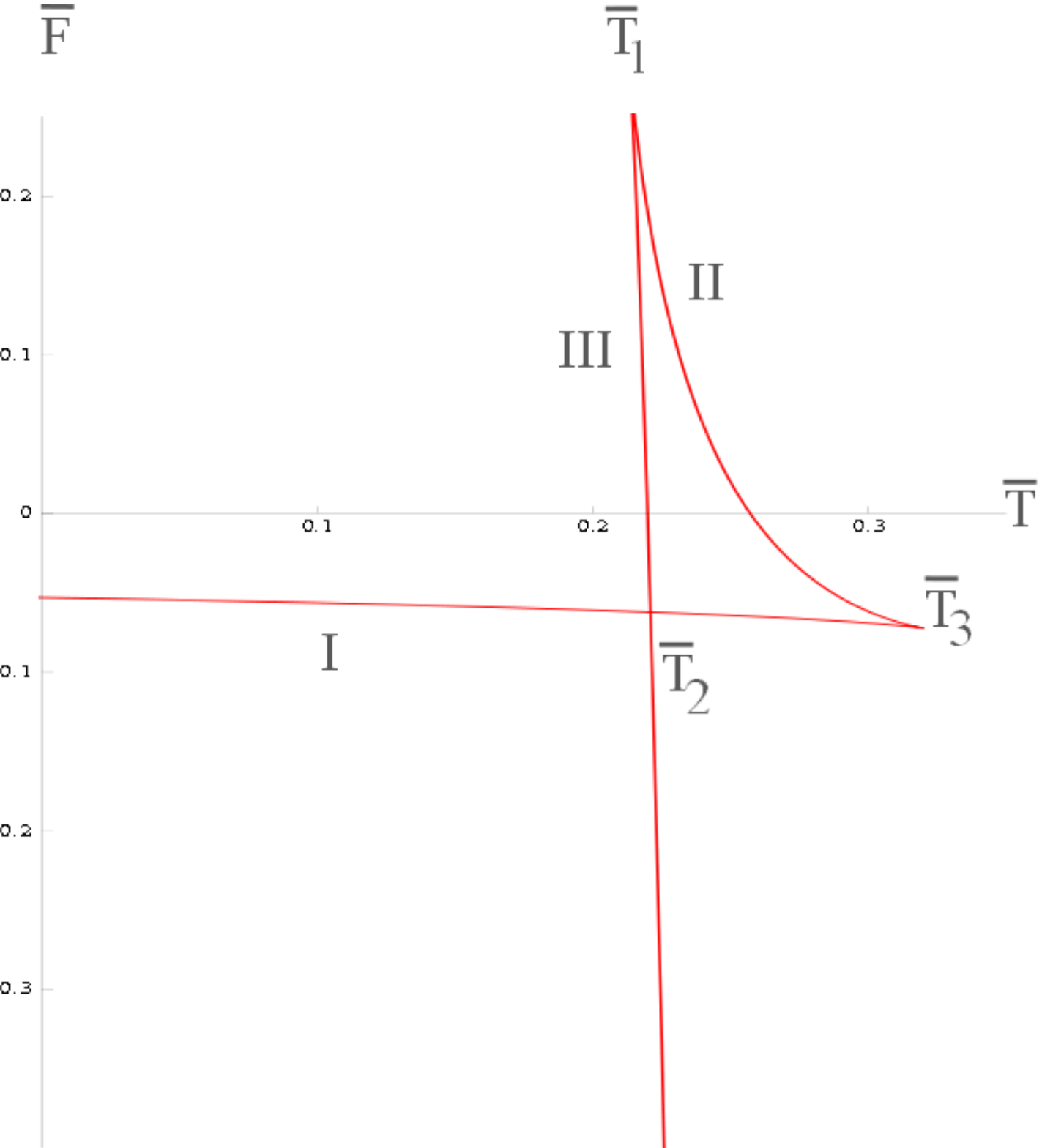}}
\quad
\subfigure[]{\includegraphics[height=4.5cm,width=6.5 cm]{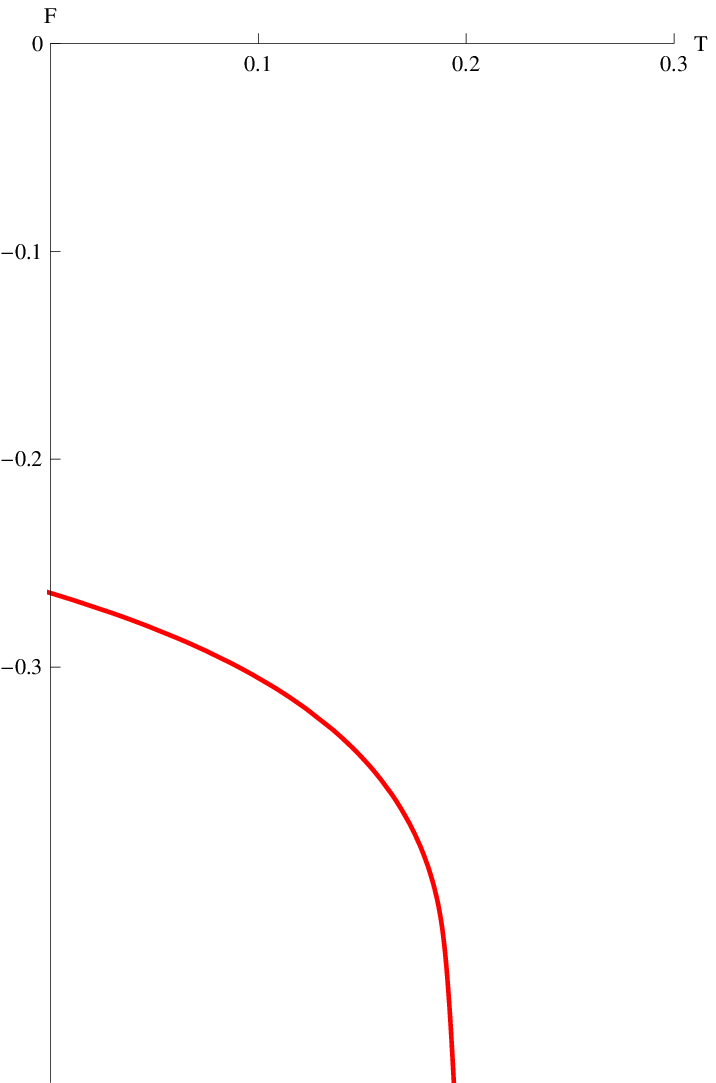}}}
\end{psfrags}
\caption{The plot (a) is for $\bar a=0.2 < \bar a_c, K=1 \,{\rm and}\, l=2$ and plot (b) is for $\bar a=0.45 > \bar a_c, K=1 \,{\rm and}\, l=2$.}
\label{F-T}
\end{center}
\end{figure}
Figure \ref{F-T} represents the plot of free energy as a function of temperature.
From the plot we take notice of the following scenario. For $\bar a > \bar a_c$, there is only one branch
with negative free energy. Thus this branch will be stable.
 For $\bar a < \bar a_c$, at low temperature free energy has only one branch (I) and as the temperature is increased, two new branches (II and III) with positive value appear at temperature
$\bar T_1$. If temperature increases further free energy of both the branches continues to decrease. Branch III cuts branch I at temperature $\bar T_2$ and becomes more and more negative at
temperature $\bar T_3$ where branch II meets branch I and both disappear. These three branches represent respectively small, intermediate and large black holes. Out of these three, the
intermediate black hole is unstable with negative specific heat while the other two are stable with positive specific heat. Bellow temperature $\bar T_1$ only branch I exist with  the free
energy less than AdS configuration. Within the range of  temperature $\bar T_1$ and $\bar T_2$, the free energy of the branch III is higher than the branch I. Thus branch I should be stable
configuration than the branch III. Once temperature crosses $\bar T_2$  the scenario is just opposite and the branch III will be stable configuration. Therefore at low temperature there is
only small black hole and once temperature increases and approaches towards $\bar T_1$  there is a nucleation of medium and large black holes occurs where as medium one is unstable.  At
$\bar T_2$  cross over from the small black hole to the large black hole takes place. Above this temperature only large black hole
exists. Therefore, it can be said that AdS configuration can not be the most stable one. Only the small or the large black hole configurations survive.

To render further support for such a scenario, we now study the Landau function. In terms of the dimensionless quantities the Landau function can be
written as follows:
\begin{equation}
\bar G=\frac{V_3 l^2 }{16\pi G_5}\big[3 \bar r^4 -4\pi T l \bar r^3 + 3 K\bar r^2- 2\bar a \bar r\big].
\end{equation}
To analyze the different phases we plot this Landau function with respect to $\bar r$  for different temperature.
\begin{figure}[h]
\begin{center}
\begin{psfrags}
\psfrag{G}[][]{$\bar G$}
\psfrag{r}[][]{$\bar r$}
\mbox{\subfigure[]{\includegraphics[width=6.5 cm]{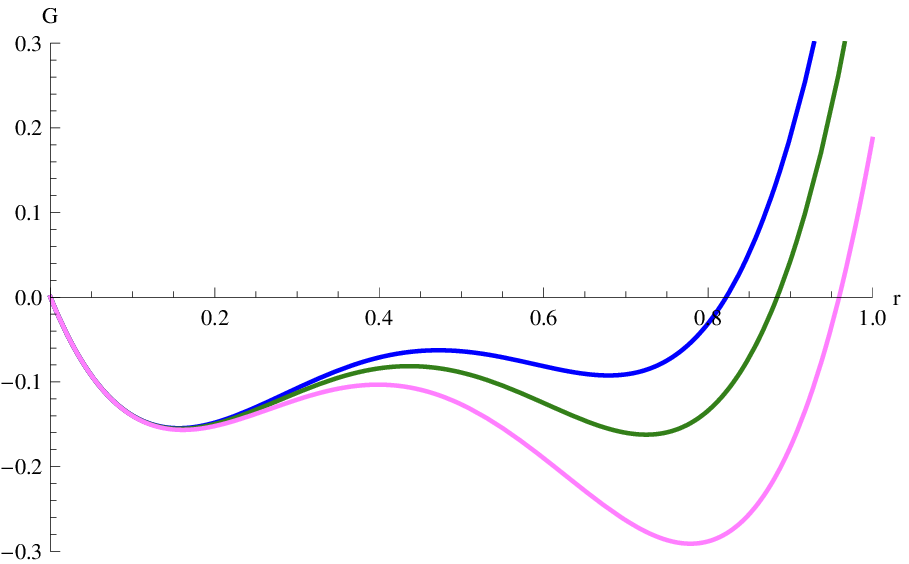}}
\quad
\subfigure[]{\includegraphics[width=6.5 cm]{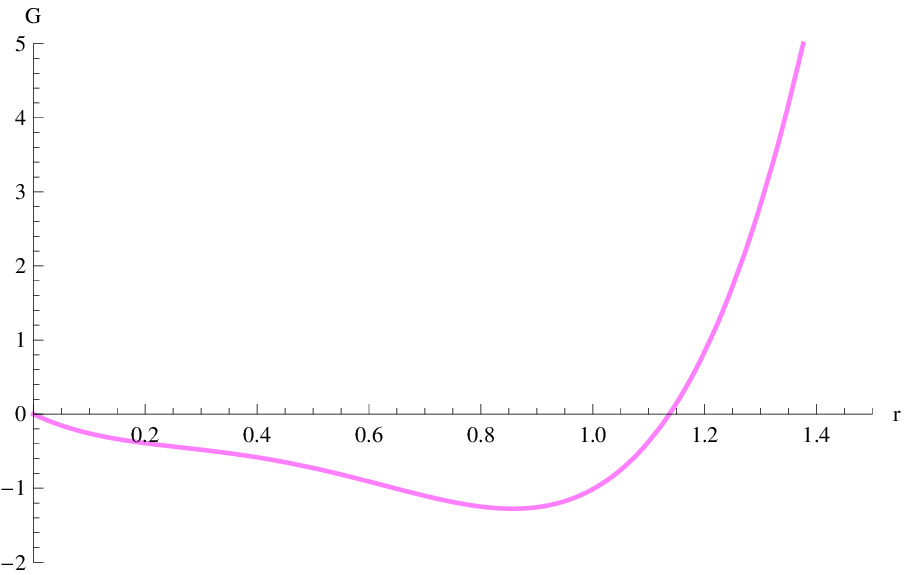}}}
\end{psfrags}
\caption{The plot (a) is for $\bar a=0.3 < \bar a_c, K=1 \,{\rm and}\, l=2$ and plot (b) is for $\bar a=0.45 > \bar a_c, K=1 \,{\rm and}\, l=2$. The blue curve corresponds to the temperature $\bar T_1  =0.208$, the green curve corresponds to $\bar T_2 = 0.21$ and the pink curve corresponds to  $\bar T_3 = 0.213$ such that $\bar T_1 < \bar T_2 < \bar T_3$.}
\label{gibbs}
\end{center}
\end{figure}
In figure \ref{gibbs}(a), for $\bar a < \bar a_c$, we plot the free energy for three different temperatures.  There exists two black hole solutions corresponding to the temperature $\bar T_1$
but the energy of the small black hole is less than the large one. So the small one will be stable. Similarly at temperature $\bar T_2$  two black hole solutions co-exist. Finally at
temperature $\bar T_3$, the energy of the large black hole is small compared to the small black hole and the large black hole will be stable configuration. Therefore we conclude that at low temperature small black hole will be the stable configuration and at temperature above a critical value large one will be stable and at the critical temperature there will be a phase transition between large and small black hole.
For  $\bar a > \bar a_c$, the Landau function has one minimum with negative value. The qualitative behaviour of the Landau function is same for different temperatures. Therefore we have only one stable black hole.

All the above calculations were done for curvature constant $K = 1$. For $K = 0 \, \rm and -1$ we find that the qualitative feature of the above thermodynamical quantities are similar to the case of $K = 1$ with $\bar a > \bar a_c$. So at any temperature, there exists a single black hole phase of finite size with non-zero entropy.

As a concluding remark the AdS space is not a stable configuration. Either small or large black hole configuration is the stable configuration. Therefore, in the dual gauge theory we may think that the bound states of a quark and anti-quark pairs do not exist. In the next section we study the dual gauge theory to check the stability of the bound state of a quark and anti-quark pairs.

\section{Dual gauge theory}\label{gaugetheory}
In the dual gauge theory we consider a probe string whose end points are attached on the boundary of the black hole background and the body of the string is hanged in to the bulk space-time. In the bulk space-time, there are two configurations of the open string. One is the U-shape configuration, where the body of the string reach up to a maximum distance from the boundary. The other one is the straight configuration where the open string reach up to the horizon of the black hole. The first configuration corresponds to the confined state of quark and anti-quark pairs and the second configuration corresponds to deconfined state of quark and anti-quark pairs. Since we are going to study the existence of bound state of quark and anti-quark pairs, so we must study the distance between the quark and anti-quark pairs living on the boundary of the black hole space time. Therefore it is convenient to move from polar coordinate to cartesian coordinate. With this aim we replace radial coordinate $r$ as $\frac{l^2}{u}$ and the metric solution of equation (\ref{genmet}) reduces to the following form,
 \begin{equation}\label{dualgrav}
ds^2= f(u)\big[-h(u)dt^2 + dx^2 + dy^2 +  dz^2 + {du^2\over h(u)}\big],
\end{equation}
\begin{equation}
f(u)= {l^2\over u^2} \quad\quad{\rm and}\quad\quad h(u)= 1+ \frac{u^2}{l^2}-{2m u^4\over l^6} - {2\over 3}{\bar a u^3\over l^3}.\nonumber
\end{equation}
In this coordinate system the boundary appears at u = 0 and the modified radius of horizon can be constructed by solving the equation,
\begin{eqnarray}
 h(u_{+})= 1-{2m u_{+}^4\over l^6} - {2\over 3}{\bar a u_{+}^3\over l^3}=0
\label{hor}
\end{eqnarray}
The large and small black hole is defined as $u_+$ is small and large respectively.

In order to study the distance between quark and anti-quark pair we start with the probe string world sheet action in the above black hole back ground. The world sheet action can be written as,
\begin{equation}
S =\int d^2\xi \mathcal{L} = \int d^2\xi \sqrt{det\; h_{\alpha \beta}} .
\end{equation}
Where the induced metric $h_{\alpha\beta} = \partial_\alpha X^\mu\partial_\beta X^\nu g_{\mu\nu}$. In this dual theory we prefer to work in the following
static gauge: $\xi^0 = t,\,\, \xi^1 = x$. For these choices the induced metric in string frame can be written as,
\begin{equation}
ds^2 = f(u)\big[-h(u)dt^2 + \big\{1 + \frac{u'^2}{h(u)}\big\}dx^2\big].
\end{equation}
Here $u'$ denotes a derivative of $u$ with respect to $x$. The Lagrangian and Hamiltonian of the quark and anti-quark pair can easily be calculated as,
\begin{equation}
\mathcal {L} = \sqrt{-det h_{\alpha \beta}} =  f(u) \sqrt{h(u) + u'^2},
\end{equation}
\begin{equation}\label{hamiltonian}
\mathcal { H} = (\frac{\partial \mathcal L}{\partial u'})u' -\mathcal {L} =  -f(u) \frac{h(u)}{\sqrt{h(u) + u'^2}}.
\end{equation}
Following these boundary conditions;
\begin{equation}
u(x = \pm \frac{L}{2}) = 0, u(x=0) = u_0 \,\,{\rm and}\,\, u'(x=0)=0,
\end{equation}
we can obtain the conserved energy of the quark and anti-quark pair as,
\begin{equation}
\mathcal {H}(x=0) = -f(u_0) \sqrt{h(u_0)}.
\end{equation}
From equation (\ref{hamiltonian}), $u'$ can also be found as,
\begin{equation}
u' = \sqrt{h(u)\Big[\frac{\sigma^2(u)}{\sigma^2(u_0)}-1\Big]},
\end{equation}
where $$\sigma(u)= f(u)\sqrt h(u).$$
Finally the distance $L$ between the quark and anti-quark pair can be calculated as,
\begin{equation}
L = \int_{-\frac{L}{2}}^{\frac{L}{2}}dx = 2 \int_0^{u_0} \frac{1}{u'}du =2 \int_0^{u_0}du \Big[ h(u)\Big\{\frac{\sigma^2(u)}{\sigma^2(u_0)}-1\Big\} \Big]^{-\frac{1}{2}},
\end{equation}
where $u_0$ is the maximum depth that the string can reach towards the black hole horizon of the background.
\begin{figure}[h]
\begin{center}
\begin{psfrags}
\psfrag{f}[][]{$L$}
\psfrag{u}[][]{$u_0$}
\mbox{\subfigure[]{\includegraphics[width=6.5 cm]{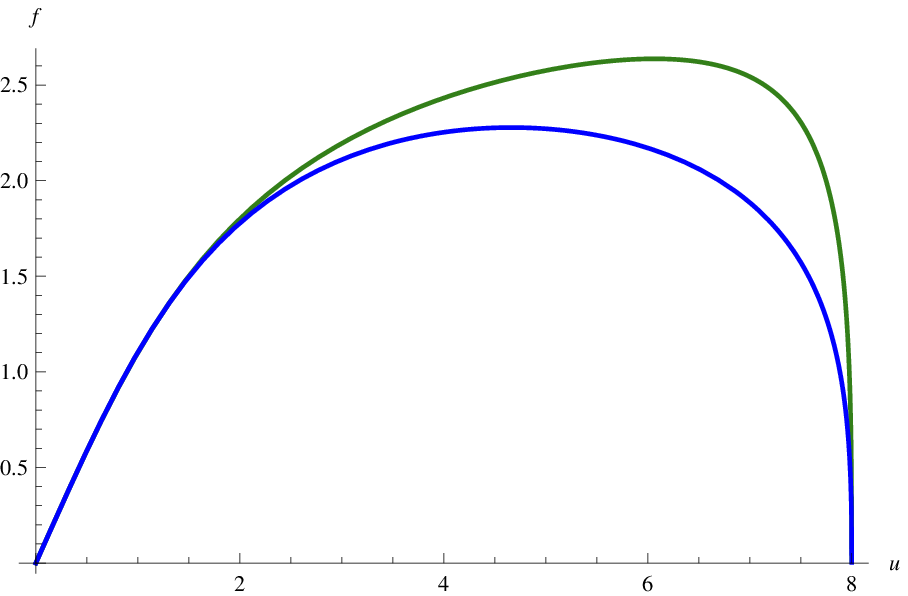}}
\quad
\subfigure[]{\includegraphics[width=6.5 cm]{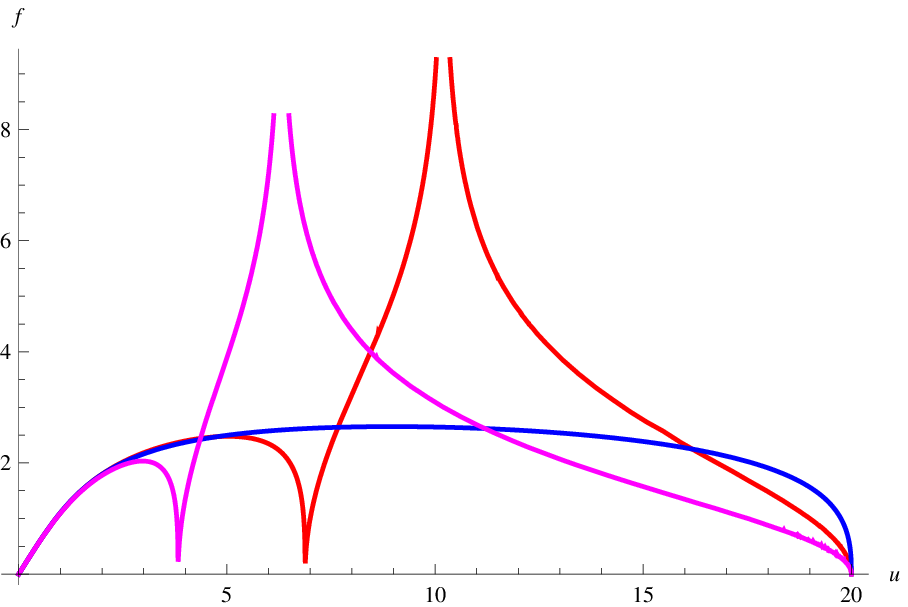}}}
\end{psfrags}
\caption{Both the plots are drawn for $K=1 \,{\rm and}\, l=1$. In plot (a) $u_+ = 8,\; \bar a=0.3(\rm green\; curve)$ and $\bar a=0(\rm blue\; curve)$. In plot (b) $u_+ = 20,\; \bar a=0.5(\rm pink\; curve),\;\bar a=0.3(\rm red\; curve)$ and $\bar a=0(\rm blue\; curve)$. }
\label{distance}
\end{center}
\end{figure}
In figure \ref{distance} we plot the distance $L$ between quark and anti-quark pair with respect to $u_0$. Notice that for large size black hole that is for $u_+$ small, irrespective of the density of the string cloud, the distance $L$ initially increases as $u_0$ approaches towards black hole horizon and finally it takes its maximum value when $u_0$ goes near to the horizon value and then breaks down to zero value when $u_0$ reaches the horizon $u_+$. However for the small black hole that is for $u_+$ large, there are two scenario. One for the string cloud density is zero, the nature of the distance $L$ is expectedly similar to the large black hole. The other one for the non zero string cloud density, the distance $L$ takes its local maximum value and then it breaks down to the zero value before reaching $u_0$ to the horizon.  Thus for any black hole back ground, an open string is in the U-shape configuration for short distance between quark and anti-quark pair and is in the straight shape configuration for large distance. Only depending on the cloud density the maximum depth $u_0$ of the probe string is changing. Therefore, for any black hole configuration, only stable deconfined phase of quark and anti-quark pairs exist in the dual gauge theory. Which is not matching with the result of \cite{Yang} since our background is different from them.

Though we have given the graph for  $\bar a < \bar a_c$ and $K= 1$; but the qualitative nature
of the graph is same for the other values of $\bar a$ and $K$. Thus we are not providing the plots for different values of $\bar a$ and $K$ and also we are not repeating the same analysis again.

\section{Summary}\label{conclude}
In this work we first study the thermodynamics of AdS-Schwarzschild black hole in presence of
external string cloud. We observe that for all values of curvature constant the black hole
configuration is stable compared to the AdS configuration. However, when the value of the  curvature constant equals to one and when the string cloud density is less than a critical value, within a certain range of temperature, there are three black
holes, while outside this range there is only one black hole. Depending on the size these
three black holes we call them as small, medium and large black holes. Among these black holes small and large one come with positive specific heats and the medium has negative one. Due to the presence of large number of strings the AdS background is deformed to a small black hole background.
In order to test their stability we study the free energy and Landau function. Finally we observe that
within the aforesaid temperature regime a  phase transition take place among the small
and large black holes which leads us to conclude that the bound state of quark and anti-quark pairs may not exist. Therefore, we study the existence of bound state of quark and anti-quark pairs in the dual gauge theory. We have shown that any black hole configuration corresponds to usual deconfined state of the gauge theory.
We therefore conclude that due to the presence of string cloud, the AdS space-time deformed to a black hole space-time. Depending on the density of the string cloud either large or small black hole background becomes stable and in the gauge theory only deconfined phase will be stable.

\section*{Acknowledgements}
 I would like to acknowledge Shankhadeep Chakrabortty, Pronita Chettri,Sudipta Mukherji and Pei-Hung Yang for going through the draft and giving their valuable comments.

\end{document}